\documentstyle[onecolumn,epsfig]{mn2e}
\def\gtwid{\mathrel{\raise.3ex\hbox{$>$\kern-.75em\lower1ex\hbox{$\sim$}}}}
\def\ltwid{\mathrel{\raise.3ex\hbox{$<$\kern-.75em\lower1ex\hbox{$\sim$}}}}
\def\\{\hfil\break}
\def\ie{{\it i.e.\ }}
\def\eg{{\it e.g.\ }}
\def\etal{{\it et al.\ }}

\newcommand{\apj}{ApJ}
\newcommand{\mnras}{MNRAS}
\newcommand{\aj}{AJ}

\begin{document}

\title{Power Spectrum Shape from Peculiar Velocity Data}
\vskip 0.5cm
\author[Watkins \& Feldman]{Richard Watkins$^{\star,1}$ \& Hume A. Feldman$^{\dagger,2}$ \\
$^\star$Department of Physics, Willamette University, Salem, OR 97301\\
$^\dagger$Department of Physics and Astronomy, University of Kansas, Lawrence, KS 66045.\\
emails: $^1$rwatkins@willamette.edu;\, $^2$feldman@ku.edu}

\maketitle
\begin{abstract}

We constrain the velocity power spectrum shape parameter $\Gamma$  in linear theory using the nine bulk--flow and shear moments estimated from four recent peculiar velocity surveys. For each survey, a likelihood function for $\Gamma$ was found after marginalizing over the power spectrum amplitude $\sigma_8\Omega_m^{0.6}$ using constraints obtained from comparisons between redshift surveys and peculiar velocity data.   In order to maximize the accuracy of our analyses, the velocity noise $\sigma_*$ was estimated directly for each survey. A statistical analysis of the differences between the values of the moments estimated from different surveys showed consistency with theoretical predictions, suggesting that all the surveys investigated reflect the same large scale flows.   The peculiar velocity surveys were combined into a composite survey yielding the constraint $\Gamma=0.13^{+0.09}_{-0.05}$. This value is lower than, but consistent with, values obtained using redshift surveys and CMB data.
\end{abstract}

\noindent{\it Subject headings}: cosmology: distance scales -- cosmology: large
scale structure of the universe -- cosmology: observation -- cosmology:
theory -- galaxies: kinematics and dynamics -- galaxies: statistics

\section{Introduction}
\label{sec-intro}

%General comments about peculiar velocities

Observations of the large scale peculiar velocity field provide an important tool for probing mass fluctuations on $\sim$ 100 h$^{-1}$ Mpc scales (h is the Hubble constant in units of 100 km s$^{-1}$ Mpc$^{-1}$). Analyses of peculiar velocity surveys used to constrain the amplitude of  mass power spectrum \cite{freudling1999,zaroubi2001} are complementary to those that employ redshift surveys alone (e.g. \cite{2df2001,tegmark2002}) or in combination with CMB data \cite{elgaroy2002,sanchez2005}. Although initially peculiar velocity surveys using redshift independent distance indicators were sparse and shallow \cite{FJ,TF} ,  large, homogeneous redshift--distance samples of galaxies and clusters have become increasingly common.   Early analyses of redshift--distance surveys \cite{aaronson,lb1988} led to the development of powerful statistical methods for the analysis of peculiar velocity data \cite{7s,Kaiser88,fw94,sw1995,wf95}, but were hindered by shallow and sparse samples. However, with the advent of larger and better samples \cite{pairwise,ph05,pp06,sarkar07} it has become increasingly clear that peculiar velocity catalogues can play an important role in the determination of cosmological parameters. 

Recently,  several studies have compared peculiar velocity surveys directly to the density field as determined from redshift surveys and put strong constraints on the combination of parameters $\sigma_8\Omega_m^{0.6}$ (for a summary, see Pike \& Hudson 2005).    These studies have involved 
peculiar velocity data derived from several different distance measures, as well as three different redshift surveys, and yet have produced remarkably consistent results.   When one considers that the constraints from these studies are also consistent with those derived from other types of studies, it seems reasonable to consider our knowledge of $\sigma_8\Omega_m^{0.6}$ to be firmly established.   

The combination of parameters $\sigma_8\Omega_m^{0.6}$ essentially determines the amplitude of the velocity power spectrum.   The shape of the power spectrum can also be constrained using peculiar velocity data, but efforts of this type have typically tried to constrain the amplitude and shape together, resulting in constraints on a two-dimensional space of parameters (see, \eg Jaffe \& Kaiser 1995, Borgani et al. 2000).   
In this {\it paper} we use peculiar velocity surveys to put constraints on power spectrum parameters directly through linear theory.    By using the constraint on $\sigma_8\Omega_m^{0.6}$ as a prior, we are able to calculate the likelihood function for the ``shape'' parameter, $\Gamma$, of the power spectrum by itself, resulting in a simple one-dimensional constraint on this parameter.

Given that our analysis uses linear theory, it is important that we properly filter out small-scale flows due to nonlinear effects that are present in peculiar velocity surveys.   For this reason, we follow Jaffe \& Kaiser (1995) in analyzing surveys using only the two lowest order moments of the velocity field, \ie the nine bulk flow and shear moments.    For the sizes and depths of the surveys that we are considering, the bulk flow and shear moments probe velocity modes with wave numbers well above the scales where nonlinear effects are thought to be important.    Significantly, the bulk flow and shear moments are also the velocity moments that have the highest signal-to-noise and thus can be determined most accurately;  indeed, little useful information is lost by discarding higher order moments.   

We apply our analysis to an extensive set of peculiar velocity surveys.   Each of these surveys employs a different method of distance estimation, and target different populations of galaxies.     We expect that each survey is affected by nonlinearities in different ways; this could be reflected as differing amounts of small-scale motion superimposed upon the large-scale linear flow reflected in the bulk flow and shear.   These small-scale motions can be quantified by a standard deviation $\sigma_*$ of the velocities remaining after the bulk flow and shear have been subtracted out.    While $\sigma_*$ is typically given a fixed value of around $300$ km/sec in this type of analysis \cite{k91,fw94},
in order to improve the accuracy of our study we have calculated the maximum likelihood value of $\sigma_*$ directly for each survey using an iterative method.   

An important question, then, is how well does the information about large-scale flows contained in these different surveys agree?    In order to answer this question quantitatively, we calculate the covariance matrix for the {\it differences} between the values of the bulk flow and shear moments for two surveys.   The covariance matrix, together with the measured differences, allows us to calculate a $\chi^2$ that reflects the probability that both surveys reflect the same underlying large-scale flow.   This probability is most useful when surveys are similar in their characteristics, as the surveys that we consider are; two surveys that probe the velocity field in different ways can have a high probability of agreement even when the values of their bulk flow and shear moments are quite different.   

In $\S$ 2, we give the details of our likelihood analysis.  In $\S$ 3 we discuss the power spectrum parametrization we use.   In $\S$ 4, we describe the peculiar velocity surveys used in the analysis.   In $\S$ 5  we present our results, and in $\S$ 6 we discuss them and compare them to  constraints on the shape parameter derived from other types of data.

\section{Likelihood Analysis}
\label{Like}

Small scale motions of galaxies reflect nonlinear evolution and can depend strongly on the specific population of galaxies that are sampled in a survey.   However, by forming velocity moments out of weighted sums of individual velocities that reflect only large scale motions, we expect small scale motions to average out so that linear theory is applicable.   
In this study, we focus on the measures of the large scale flow given by the first and second order moments of a taylor expansion of the velocity field, 
\begin{equation}
{\bf v_i}({\bf r}) = {\bf u_i} + p_{ij}{\bf r_j} + \cdots ,
\end{equation}
where ${\bf u}$ is the bulk flow vector and $p_{ij}$ is the shear tensor, which can be taken to be symmetric.   
While the interpretation of these measures depend on the specific distribution of galaxies in a survey as well as measurement errors,  the surveys we consider are similar enough that they form a good basis for comparison \cite{sarkar07}.   

The samples that we  consider consist of a set of $N$ galaxies, each with a position vector ${\bf r_n}$ and a measured line-of-sight peculiar velocity $S_n={\bf v}\cdot {\bf\hat r_n}$ with individual measurement error $\sigma_n$.   Following \cite{k91} , we group the bulk flow and shear components into a 9 component vector $a_p$, so that the galaxy velocities can be modeled as 
\begin{equation}
S_n =  a_pg_p({\bf r_n}) + \delta_n,
\label{eq:gal-vel}
\end{equation}
where the 9 component vector $g_p({\bf r}) = (\hat r_x,\hat r_y,\hat r_z,r \hat r_x\hat r_x,$ $r\hat r_x\hat r_y
,r\hat r_x\hat r_z,r\hat r_y\hat r_y,r\hat r_y\hat r_z,r\hat r_z\hat r_z)$.   In what follows we take our coordinate axes to correspond to galactic coordinates.    
The deviation from the model, $\delta_n$, contains contributions from both small scale motions and measurement errors.    We assume that the $\delta_n$ values are gaussian distributed and have a variance given by $\sigma_*^2 + \sigma_n^2$ \cite{Kaiser88}.   Here $\sigma_*$ represents the small-scale linear and nonlinear motions that are not accounted for by the bulk flow and shear.   We shall refer to $\sigma_*$ as the velocity noise.   

Under these assumptions, the likelihood function for the moments is
\begin{equation}
{\cal L}(a_p;\sigma_*) = \prod_n {1\over \sqrt{\sigma_n^2+\sigma_*^2}}\exp \left(-{1\over 2}{\left[
S_n - a_pg_p({\bf r}_n)\right]^2\over 
\sigma_n^2+\sigma_*^2}\right).
\label{eq:ML}
\end{equation}

For a given survey, we would like to find the maximum likelihood values for the moments $a_p$ and for $\sigma_*$.    In previous analyses of this type, the value of $\sigma_*$ has sometimes been determined by an educated guess (see, \eg \cite{Kaiser88,fw94,jk95,ph05}). 
However,  we note that since surveys generally sample a specific population of galaxies, each of which can be differently  affected by nonlinear overdensities, we expect that each survey has, in principle, a different value for $\sigma_*$.    Thus, here we  determine the value of $\sigma_*$ directly from each survey.   We find the maximum likelihood values for the moments $a_p$ and $\sigma_*$ using an iterative method.    
We start by making a ``guess" of $\sigma_*=300$ km/sec.    
Treating $\sigma_*$ for the moment as constant, the maximum likelihood solution for the $a_p$ is given by
\begin{equation}
a_p = A_{pq}^{-1}\sum_n {g_q({\bf r}_n) S_n\over  \sigma_n^2+\sigma_*^2},
\label{eq:MLa}
\end{equation}
where
\begin{equation}
A_{pq} = \sum_n {g_p({\bf r}_n)g_q({\bf r}_n)\over  \sigma_n^2+\sigma_*^2}.
\end{equation}

Using these estimates in Eq. (\ref{eq:ML}) and treating them as constant, we then find the maximum likelihood value of $\sigma_*$, which  now replaces our original guess and can be used to calculate a refined set of maximum likelihood values for the $a_p$.   This process is repeated until the estimates converge, which in practice only takes two or three iterations.    

In order to compare our estimates of the moments $a_p$ to the expectations of theoretical models, we calculate the covariance matrix, which from Eq. (\ref{eq:MLa}) can be written as
\begin{equation}
R_{pq}=\langle a_p a_q\rangle = A^{-1}_{pl}A^{-1}_{qs}\sum_{n,m} {g_l({\bf r}_n)g_s({\bf r}_m)\over
(\sigma_n^2+\sigma_*^2)(\sigma_m^2+\sigma_*^2)}\langle S_nS_m\rangle,
\label{eq:cov}
\end{equation}
where $\langle S_nS_m\rangle$ can be written in terms of the linear velocity field ${\bf v}({\bf r})$ 
and the variance of the scatter about it,
\begin{equation}
\langle S_nS_m\rangle = \langle {\bf \hat r}_n\cdot {\bf v}({\bf r}_n)\ \   {\bf\hat r}_m\cdot {\bf v}({\bf r}_m)\rangle
+ \delta_{nm}(\sigma_*^2 + \sigma_n^2).
\label{eq:galv}
\end{equation}
 Plugging Eq. (\ref{eq:galv}) into Eq. (\ref{eq:cov}),  the covariance matrix reduces to two terms, 
 \begin{equation}
 R_{pq} = R^{(v)}_{pq} +  R^{(\epsilon)}_{pq} . 
 \end{equation}
 The second term, called the "noise" term, can be shown to be
 \begin{equation}
 R^{(\epsilon)}_{pq}  = A^{-1}_{pq}.
 \end{equation}
 The 
 first term is given as an integral over the matter fluctuation power spectrum, $P(k)$, 
 \begin{equation} 
 R^{(v)}_{pq}  = {\Omega_{m}^{1.2}\over 2\pi^2}\int_0^\infty
 dk \ \  {\cal W}^2_{pq}(k)P(k),
 \label{eq:covv}
 \end{equation}
 where the angle-averaged tensor window function is 
 \begin{equation}
 {\cal W}^2_{pq} (k)=A^{-1}_{pl}A^{-1}_{qs} \sum_{n,m} {g_l({\bf r}_n)g_s({\bf r}_m)\over
(\sigma_n^2+\sigma_*^2)(\sigma_m^2+\sigma_*^2)}
	\int {d^2{\hat k}\over 4\pi}\ \left({\bf \hat r}_n\cdot {\bf \hat k}\ \ {\bf \hat r}_m\cdot {\bf \hat k}\right) 
	\exp\left[i{\bf k}\cdot ({\bf r}_n- {\bf r}_m)\right].
\end{equation}

Given a peculiar velocity survey and the values of its 9 moments, we can write the likelihood of a theoretical model used to calculate the covariance matrix as 
\begin{equation}
{\cal L} =  {1\over |R|^{1/2}}\exp\left(-{1\over 2}a_pR^{-1}_{pq}a_q\right).
\label{eq:likelihood}
\end{equation}
As we describe below, we  use this equation in order to place constraints on the parameters of cosmological models; in particular  $\Gamma$, the parameter that determines the shape of the power spectrum.

Both the bulk flow and shear moments probe primarily large-scale motions and thus have window functions that are peaked at large scales.   However, our prior constraint on $\sigma_8\Omega_m^{0.6}$ essentially fixes the amplitude of the power spectrum on relatively small scales.   Since the shape parameter $\Gamma$ controls the relative distribution of power between large and small scales (smaller $\Gamma$ corresponds to relatively more power on large scales),  increasing $\Gamma$ generally results in smaller values for the elements $R_{pq}$ of the covariance matrix.    Thus Eq. (\ref{eq:likelihood}) shows that surveys with large values of the bulk flow and shear moments generally favor smaller values of $\Gamma$.

The window function for a given survey carries information about how the moments of that survey sample the power spectrum.   Since each survey has a unique window function, the values of the moments are not strictly comparable between surveys \cite{wf95}.    However, since the volumes occupied by the surveys we  consider overlap strongly, we expect the values of the moments of these surveys to be highly correlated.   In order to quantify the agreement between two different surveys, say, survey $A$ and survey $B$, we  use the covariance matrix for the {\it difference} between the values $a_p^A$ and $a_p^B$ of the moments for the two surveys,
\begin{equation}
R^{A-B}_{pq}=  \langle (a_p^A - a_p^B)(a_q^A - a_q^B)\rangle= R^A_{pq} + R^B_{pq}-R^{AB}_{pq}-R^{AB}_{qp},
 \end{equation}
where the cross-terms are given by
\begin{equation} 
 R^{AB}_{pq}  = {\Omega_{m}^{1.2}\over 2\pi^2}\int_0^\infty
 dk \ \ ({\cal W}^{AB})^2_{pq}(k)P(k),
\label{eq:covv-diff}
\end{equation}
and
 \begin{eqnarray}
({\cal W}^{AB})^2_{pq} (k)=
 (A^A)^{-1}_{pl}(A^B)^{-1}_{qs}
 	 \sum_{n,m} {g_l({\bf r}^A_n)g_s({\bf r}^B_m)\over
	((\sigma^A_n)^2+(\sigma^A_*)^2)((\sigma^B_m)^2+(\sigma^B_*)^2)} \hspace{1.5cm}\nonumber \\
	\times \int {d^2{\hat k}\over 4\pi}\ \left({\bf \hat r}^A_n\cdot {\bf \hat k}\ \ {\bf \hat r}^B_m\cdot {\bf \hat k}\right) 
		\exp\left[i{\bf k}\cdot ({\bf r}^A_n- {\bf r}^B_m)\right]. \nonumber\\
\end{eqnarray}

Note that we have assumed here that the nonlinear contributions to the velocities of the galaxies represented by $\sigma_*$ in the two surveys are uncorrelated.    This is not likely to be true in reality, since galaxies in the same local neighborhood are affected by the same small-scale flows.   However, this will always cause us to underestimate the expected amount of correlation between surveys.   Thus our results on how well two surveys agree should be considered as upper bounds.   

\section{Power Spectrum Parameters}
\label{model}

As shown in Eqs. (\ref{eq:covv}) and (\ref{eq:covv-diff}), in linear theory the variances for the velocity components $a_p$ are given by integrals over the power spectrum $P(k)$ multiplied by the factor $\Omega_m^{1.2}$.   The power spectrum $P(k)$ can itself be modeled as an initial power law $k^n$, where we  assume the usual $n\approx 1$, times the square of the transfer function $T(k)$, so that $P(k)\propto kT^2(k)$.   We  set the overall amplitude of the power spectrum in the usual way through the constant $\sigma_8$, the amplitude of matter density perturbations on the scale of $8\ h^{-1}$Mpc.   
The transfer function is generally parametrized in terms of the ``shape" parameter $\Gamma \approx\Omega_m h$.    We  follow Eisenstein and Hu (1998) in writing the transfer function as
\begin{eqnarray}
T(k) &=& {L_o\over L_o+ C_o(k/\Gamma)^2}\nonumber , \\
\label{eq:PS}
L_o(k) &=& \ln (2e + 1.8(k/\Gamma), \\
C_o(k) &=& 14.2 + {731\over 1 + 62.5(k/\Gamma)}\nonumber ,
\end{eqnarray}
which is valid in the zero-baryon approximation.   While Eisenstein and Hu (1998) include a more accurate model
for the power spectrum that includes the effects of baryons,  this model depends on several more parameters and does not have a constant $\Gamma$.   We have found that there is not enough information in the peculiar velocity data to break the degeneracies between these parameters, and that a two parameter model is sufficient given the precision of our results.   We will revisit the issue of baryons in the discussion section ($\S$ 6) when we interpret our results.

\section{Peculiar Velocity Surveys}
\label{surveys}

 We have applied our analysis to four  peculiar velocity catalogues.   These catalogues vary in their sample size, depth, distance measurement method, and typical measurement errors.   Our subset of the spiral field {\it I}-band (SFI) survey (Haynes et al. 1999a, 1999b)  consists of  1016 galaxies  at distances $d < 60 $\  $h^{-1}$Mpc with velocities $|v| < 2000$ km/sec.   The distance measurements, obtained using the {\it I}-band Tully-Fisher (TF) relation, have been corrected for Malmquist and other biases as described in Giovanelli \etal (1997).   The distance cut was made to avoid the bias toward inwardly moving galaxies near the redshift limit of the survey.   The cut on large velocity galaxies was made to prevent them from having an undue effect on our results.   We have tested that our results are not sensitive to the precise way in which these cuts are made.   Distance errors on this sample are of order 16\%.

  Our subset of the Nearby Early-type Galaxy (ENEAR) survey  (da Costa et al. 2000a, 2000b) 
contains 535 galaxies at distances $d<70$ $h^{-1}$Mpc with velocities $|v| < 2400$ km/sec.    The $D_n-\sigma$ distance estimates  in this catalogue have been corrected for inhomogeneous Malmquist Bias \cite{Bernardi02}.   A correction has also been applied for the bias toward inwardly moving galaxies at the redshift limit of the catalogue; however, given the large size of this correction for galaxies near the limit, we chose to apply a distance cut in order to avoid the effects of this bias and its correction on our sample.     Distance errors for this sample are of the order of $18\%$ for individual galaxies with groups of size $N$ having their distance errors reduced by a factor of $\sqrt{N}$.   

Our subset of the $I$-band surface brightness fluctuation survey of Galaxy distances (SBF; \cite{sbf})  contains 257 E, S0, and early-type spiral galaxies out to a redshift of about $4000$ km/sec.   We have removed galaxies with percentage distance errors ($>20\%$) as well as three high-velocity galaxies with redshifts $> 4000$ km/sec.     More than half of the galaxies in our sample have distance errors below $10\%$, with the smallest errors being less than $5\%$.    With errors of this magnitude, our sample is immune from the effects of Malmquist bias \cite{sbf2}.

Finally, we have included a small sample of 73 supernovae of Type Ia (SNIa;  Tonry et al. 2003) whose redshifts extend to 8000 km/sec.   Typical Distance errors for this sample are about $7\%$; one supernova whose distance error exceeded $20\%$ was not included.   As with SBF, this sample should be immune from Malmquist Bias.

\section{Results}
\label{results}

The first step in our analysis is to obtain estimates for the moments $a_p$ and the velocity noise $\sigma_*$ for each catalogue by maximizing the likelihood function (Eq. \ref{eq:likelihood}) as described above.   The calculated moments and $\sigma_*$ for each survey are given in Table 1.   
Note that the estimates for the bulk flow vectors are quite similar, as one would expect for surveys that occupy similar volumes.   The shear moments are somewhat less consistent due to both the facts that these moments are estimated less accurately and also depend more strongly on the details of the sample geometries.    The estimates of $\sigma_*$ are all of the expected magnitude.    We shall leave a more detailed examination of the $\sigma_*$ estimates for the discussion section.

\vspace{0.5cm}
\begin{center} 
\parbox{6in}{\small {\bf Table 1:} Maximum likelihood estimates for the values of the three bulk flow and six shear moments as described in Section~\ref{Like} (equation \ref{eq:gal-vel}), and the velocity noise $\sigma_*$, for each of the surveys of interest and the composite survey.   }
\end{center} 
\begin{center} 
\begin{tabular}{l|rrr|rrrrrr|c} 
\hline 
Survey 	& \multicolumn{3}{|c|}{Bulk Flow} & \multicolumn{6}{|c|}{Shear}      & $\sigma_*$\\
		& \multicolumn{3}{|c|}{(km s$^{-1}$)}		&\multicolumn{6}{|c|}{(km s$^{-1}$Mpc$^{-1}$)}  & (km s$^{-1}$) \\ \hline
   SFI	       	&       69.8 &        -183 &        47.6 &        1.35 &       -2.94 &        3.37 &       -3.08 &        -3.2 &          -7 &         413\\ 
ENEAR	&        142 &        -204 &          47 &        3.65 &       -2.39 &        5.11 &       -4.23 &        -4.3 &       -3.86 &         386\\ 
       SBF   	&        120 &        -309 &         206 &        9.88 &       -3.45 &       -2.99 &        3.41 &       -1.99 &         -16 &         304\\ 
  SNIa      	&      57.6 &        -419 &        40.7 &        3.39 &       -3.25 &        5.19 &        2.62 &      -0.848 &       -3.14 &         327\\ 
Composite&        98.4 &        -216 &        90.6 &         2.6 &       -3.27 &        3.58 &       -2.95 &       -3.84 &       -6.44 &         \\  \hline 
\end{tabular} 
\end{center} 
\vspace{0.7cm}

\begin{figure}
     \includegraphics[width=15cm]{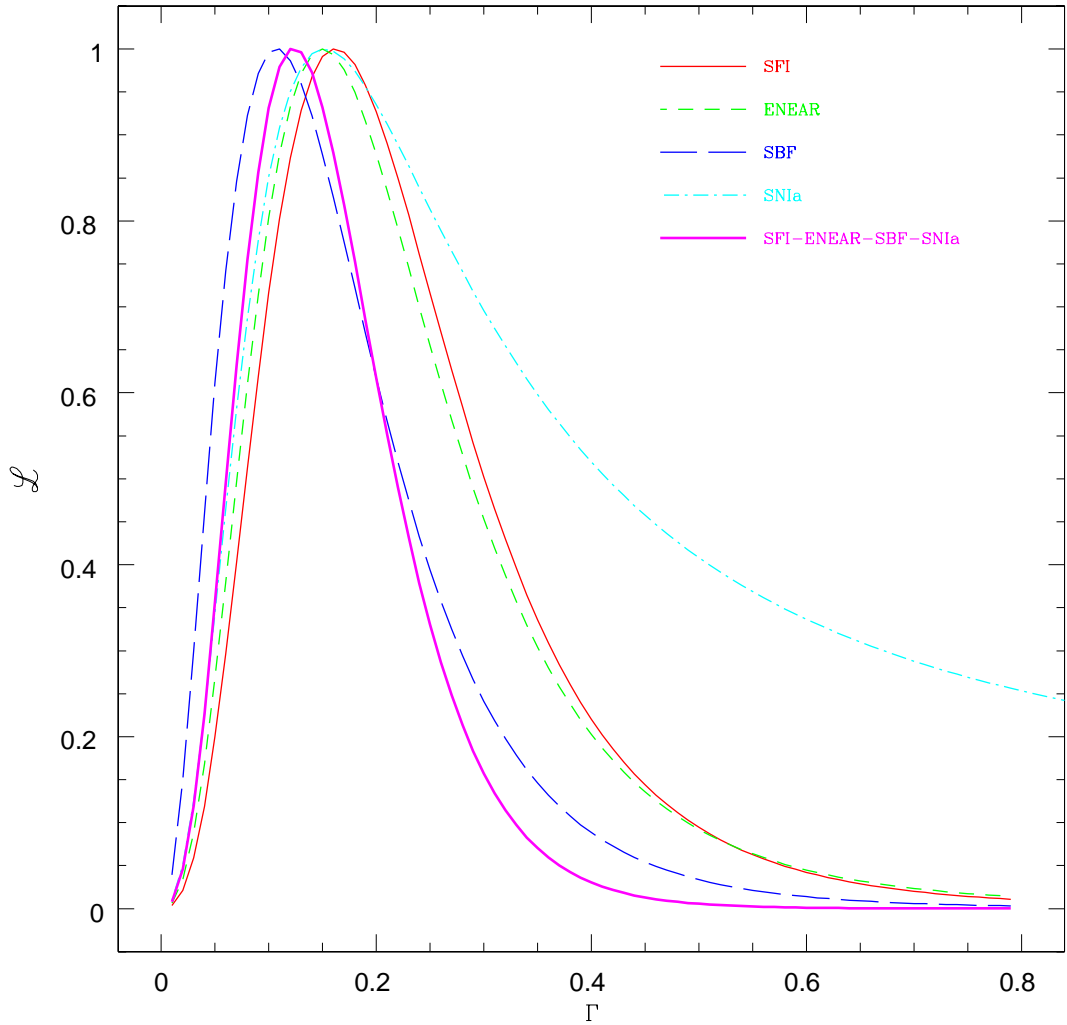}
        \caption{The likelihood functions for $\Gamma$ obtained from each of the surveys. The surveys give consistent results.   The thick solid line is the likelihood function for the composite survey.}
        \label{fig:likelihood}
\end{figure}

Next, we use Eq. (\ref{eq:likelihood}) to calculate likelihoods for model parameters for each catalogue.
From Eq. (\ref{eq:PS}), we can see that in the context of our model, the theoretical covariance matrix for the velocity moments is completely specified by two parameters; the amplitude, given by $\sigma_8\Omega_m^{0.6}$, and the shape parameter $\Gamma$.    As was mentioned above, the amplitude parameter has been strongly constrained;  in particular,  by comparisons of peculiar velocity data and redshift surveys.   Pike and Hudson (2005) summarize these constraints and combine several of them to obtain $\sigma_8(\Omega_m/0.3)^{0.6}= 0.85\pm 0.05$.    However, we feel that the averaging done on correlated data is not entirely justified, and that the constraint $\sigma_8(\Omega_m/0.3)^{0.6}= 0.84\pm 0.1$,  or $\sigma_8\Omega_m^{0.6}=0.41\pm 0.05$, which coincides with that obtained by Zaroubi et al. (2002), is more representative of the strength of the constraint that can be placed on the amplitude using peculiar velocity and redshift data.    This constraint is also consistent with measurements of this combination of parameters using other methods (for a discussion see Pike and Hudson (2005)).   Rather than treating the amplitude as a free parameter, we instead chose to adopt this constraint  as a prior and to marginalize over it.    Thus, given a peculiar velocity catalogue,  we are able to calculate the likelihood function for the single parameter $\Gamma$ which determines the shape of the power spectrum.    Note that although the amplitude and shape of the power spectrum are theoretically related through a common dependence on $\Omega_m$; rather than presuppose this dependence we  choose to treat them as independent parameters.   We  revisit the interdependence of these parameters in the discussion section.

In Figure \ref{fig:likelihood} we plot the likelihood functions obtained from each of the surveys.   These likelihood functions are asymmetric and have nongaussian tails.   In Table 2 we give the maximum likelihood values of $\Gamma$ for each survey together with the region around the maximum that contains $68\%$ of the probability under the curve.   We also list the $\chi^2$ at the maximum likelihood, where 
\begin{equation}
\chi^2 = \sum_{p,q} a_pR^{-1}_{pq}a_q.
\end{equation}
These values show that the peculiar velocity data is quite consistent with the power spectrum model we are considering.   

\vspace{0.5cm}
\begin{center} 
\parbox{2.2in}{\small {\bf Table 2:} The maximum likelihood value of $\Gamma$ and the $\chi^2$ for 9 degrees of freedom for each of the surveys.  The last entry is the results for the composite survey.}
\end{center} 
\begin{center} 
\begin{tabular}{l|c|c|c} 
\hline
Survey	& \quad\(\Gamma\) 		  & $\chi^2$\\ \hline
SFI 		&   $0.16^{+0.13}_{-0.07}$ &  10.70\\ 
ENEAR 	&   $0.15^{+0.13}_{-0.07}$ &  10.03\\
SBF 		&   $0.11^{+0.10}_{-0.06}$ &   7.42\\ 
SNIa 	&   $0.15^{+0.90}_{-0.11}$ &   7.62\\ 
Composite&   $0.13^{+0.09}_{-0.05}$ &  10.41\\ 
\hline
\end{tabular}
\end{center} 
\vspace{0.7cm}

The fact that all of the surveys that we have considered yield consistent results for the value of $\Gamma$ does not necessarily indicate consistency in the actual values of their moments.    
In order to check for more detailed compatibility between the surveys we consider the question of whether the {\it differences} between the values of the individual  moments of any two surveys, $a_p^A - a_p^B$, are consistent with that predicted by the theoretical models, \ie are the measurement errors, the velocity noise, and the differences in how each survey probes the power spectrum large enough to explain the differences in the moments.
 Again, we use a $\chi^2$ analysis;   calculating  the covariance matrix $R^{A-B}_{pq}$ as described above we form
\begin{equation}
\chi^2 = \sum_{p,q} (a^A_p-a^B_p)(R^{A-B}_{pq})^{-1}(a^A_q-a^B_q).
\end{equation}
It turns out that the $\chi^2$ calculated in this way does not depend very strongly on  $\Gamma$ in the region of interest.   For simplicity, then, we report $\chi^2$ values calculated for the single value of $\Gamma = 0.14$.
Other values of $\Gamma$ give similar results.

The results of this analysis are tabulated in Table 3.  They show good consistency between the catalogues for the favored range of $\Gamma$ values.    Thus the velocity moments of the surveys that we consider agree not only in magnitude, but also in value, inasmuch as they are expected to given measurement errors, velocity noise,  and differences in survey volumes.    

\vspace{0.5cm}
\begin{center} 
\parbox{1.7in}{\small {\bf Table 3:} $\chi^2$ for 9 degrees of freedom for the differences between the surveys.}
\end{center} 
\begin{center} 
\begin{tabular}{l|r} 
\hline 
Surveys		& $\chi^2$ \\ \hline
SFI--ENEAR 	& 4.606      \\                  
SFI--SBF 		& 7.430  \\                  
SFI--SNIa 		& 6.104   \\                  
ENEAR--SBF 	& 8.306    \\                   
ENEAR--SNIa 	& 3.299     \\                   
SBF--SNIa 	& 5.194 \\  
\hline                 
\label{table:diff}
\end{tabular} 
\end{center} 
\vspace{0.5cm}

Given our result that the four surveys that we have studied are consistent with each other, it seems reasonable to combine them into a composite survey which can then be used to obtain the strongest possible constraint on $\Gamma$.    Since the different values of $\sigma_*$ for the various surveys  reflect differences in the populations and distance measures between the surveys, we assign each galaxy in the composite survey the value of $\sigma_*$ of its parent survey.    In Figure \ref{fig:likelihood}, we show the likelihood function for $\Gamma$ resulting from the composite surveys, with the maximum likelihood value being $0.13^{+0.09}_{-0.05}$ .    In Table 1 we give the maximum likelihood values for the bulk flow and shear moments for the composite survey.  In Table 2 we present the maximum likelihood value of $\Gamma$ for the composite survey together with its associated $\chi^2$.

\section{Discussion}

Our results show consistency between catalogues containing galaxies of different morphologies and using different methods for determining velocities, confirming that each of these catalogues trace the same large-scale velocity field within uncertainties.    While previous studies have shown consistency in the bulk flow vectors calculated from different surveys \cite{hudson99,sarkar07}, our result is the first to directly compare both bulk flow and shear moments.  

The samples we considered do show differences on small scales; in particular, the velocity dispersion about the bulk flow and shear motions, represented by $\sigma_*$,  shows a range of values.   
While these differences in $\sigma_*$ could arise from how different galaxy populations respond to nonlinearities,  underestimates of measurement errors are also a likely source for $\sigma_*$.        

Although we have used a two-parameter model of the power spectrum that is strictly valid only for the zero-baryon case, where theoretically $\Gamma= \Omega_m h$, in interpreting our results for $\Gamma$ it is possible to include the effects of baryons to a first approximation.   In particular, Sugiyama (1995) has determined that $\Gamma$ scales with baryonic density $\Omega_b$ as
\begin{equation}
\Gamma = \Omega_m h \exp \left[ -\Omega_b\left(1+\sqrt{2h}/\Omega_m\right)\right].
\label{eq:sug}
\end{equation}
The parameters in this formula are tightly constrained by microwave background studies \cite{wmap1};
specifically, $h=0.732^{+0.031}_{-0.032}$, $\Omega_m=0.241\pm0.034$, and $\Omega_b h^2 = 0.0223^{+0.00075}_{-0.00073}$, which can be combined to give $\Omega_b=0.0416\pm0.0049$. Plugging these values into Eq. (\ref{eq:sug}) and propagating uncertainties gives the result $\Gamma = 0.137\pm0.025$, which is clearly consistent with our results.   While this model is not as accurate as that of Eisenstein \& Hu (1998),  the latter introduces complications in interpretation due to its $\Gamma$ having $k$ dependence.   We have found that the use of the more complicated model in our analysis did not change our results significantly given the precision that can be achieved using the available data.       

The constraint that we have obtained is also consistent with other measurements of the power spectrum.   In particular,  an analysis of the 2dF galaxy redshift survey has found that $\Gamma = 0.168\pm 0.016$ \cite{cole2005} for blue galaxies, whereas in earlier measurement \cite{2df2001} they found $\Gamma=0.20\pm0.03$. The SDSS collaboration found $\Gamma=0.213\pm0.023$ for all galaxies \cite{tegmark2004} and $\Gamma=0.207\pm0.030$ for luminous red galaxies (LRG) \cite{pope2004}.  Using CMB data from the WMAP experiment  together with the SDSS LRG data set H\"utsi (2006) found $\Gamma= 0.202^{+0.034}_{-0.031}$.   

\begin{figure}
     \includegraphics[width=15cm]{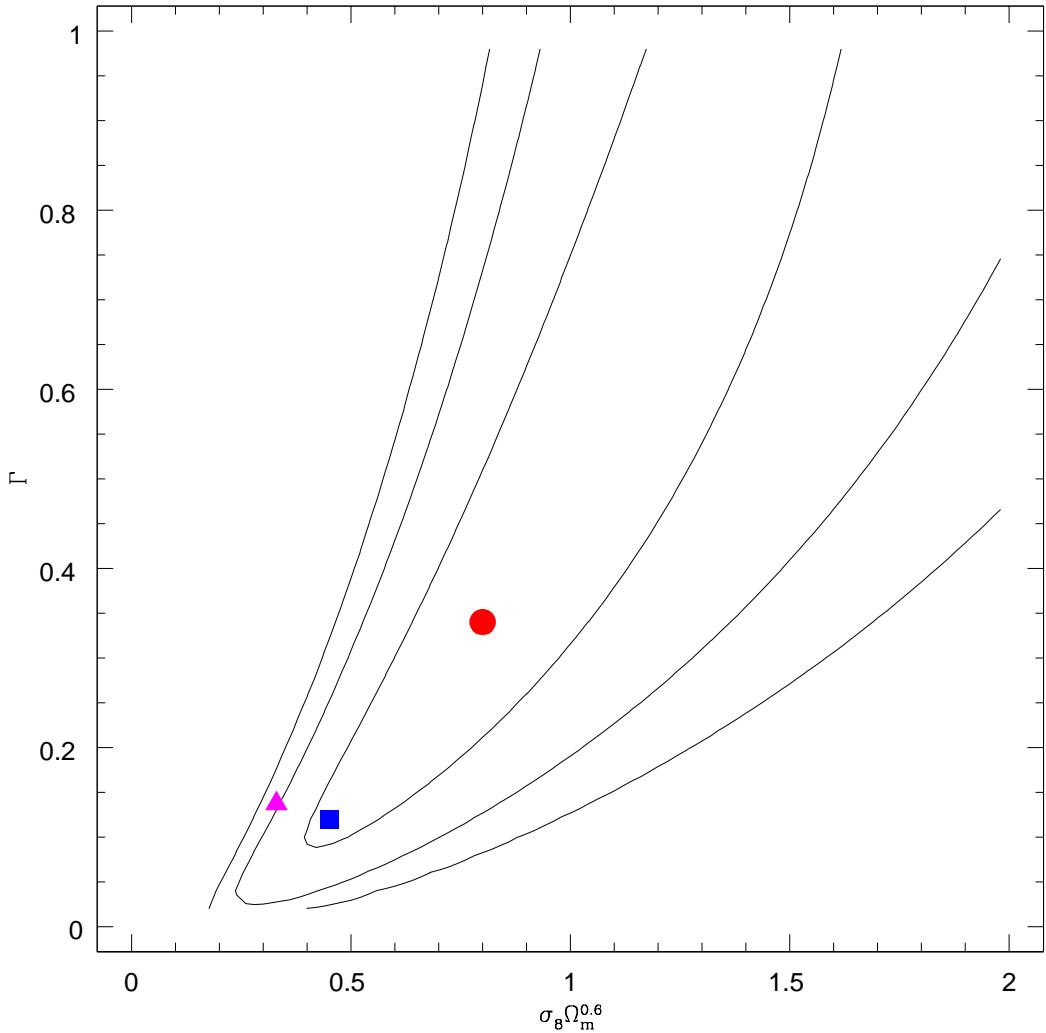}
       \caption{The 2-dimensional likelihood contours for the composite survey. The dot indicates the maximum likelihood.  The square represents the maximum likelihood value for $\Gamma$ when marginalizing over $\sigma_8\Omega_m^{0.6}$  as described in the text. The triangle shows the central values from WMAP as shown in the text. The contour levels contain 68\%, 95\% and 99\% of the total likelihood in the region of parameter space shown in the plot.   Note that the location of these contours would change if a different  region of parameter space was selected.}
         \label{fig:cont}
\end{figure}

We have used as a prior the constraint on $\sigma_8\Omega_m^{0.6}$ derived from comparisons of peculiar velocity and redshift surveys, which is appropriate for an analysis of velocity data.   However, other measurements of this quantity exist, in particular from microwave background studies \cite{wmap1}.   Using the value of $\Omega_m$ from above, together with the WMAP measurement of $\sigma_8 = 0.776^{+0.031}_{-0.032}$, and propagating errors, we find that $\sigma_8\Omega_m^{0.6}=0.330^{+0.041}_{-0.042}$, which is about one standard deviation below the central value we have used in this paper.   In order to understand the effect of the prior for $\sigma_8\Omega_m^{0.6}$, we show in Fig.~\ref{fig:cont} the likelihood contours for the composite survey over the full 2-dimensional parameter space of $\Gamma$ and $\sigma_8\Omega_m^{0.6}$ .   From the plot it is clear that using a lower the value of $\sigma_8\Omega_m^{0.6}$ would lead to a maximum likelihood value for $\Gamma$ that is smaller, although not significantly so.     The plot also illustrates the substantial degeneracy between the two parameters; the ridge that extends diagonally across contains a large region where the likelihood is close to its maximum.   Without applying a prior, as we have done, it is difficult to obtain meaningful constraints from the full 2-dimensional analysis.

It is noteworthy that combining all of the survey data into a single survey did not improve the constraint on $\Gamma$ as much as one might have expected.   While combining the data did improve the measurements of the bulk flow and shear, cosmic variance limits how strong a constraint that even perfect knowledge of these nine quantities measured for a single volume can provide.   It is important to remember that the power spectrum determines only the {\it variances} of these nine quantities;   thus our situation is similar to that of trying to constrain the variance of a statistical distribution from nine numbers drawn from that distribution.   In light of this, significant improvements of our constraint will be difficult to obtain.    Increasing the number of moments  by expanding the velocity field to higher order is likely to move the analysis into a regime where linear theory is no longer applicable.    Deeper surveys, whose bulk flow and shear are sensitive to larger scale power, could strengthen our constraint; however, given that measurement errors typically increase linearly with distance, the number of galaxies required for a reasonable analysis is prohibitive.

\noindent{\bf Acknowlegements:} 
RW acknowledges the support of an Atkinson Faculty Grant from Willamette University.   
HAF has been supported in part by a grant from the Research Corporation and by the University of Kansas General Research Fund (KUGRF). We would like to thank Riccardo Giovanelli and the SFI team and Gary Wegner and the ENEAR team for providing us with their catalogues.

\end{document}